\documentclass[fleqn,usenatbib]{mnras}
\usepackage{newtxtext,newtxmath}
\usepackage[T1]{fontenc}
\usepackage{ae,aecompl}
\usepackage{graphicx}
\usepackage{amsmath}
\usepackage{amssymb}
\usepackage{xcolor}
\usepackage{comment}
\usepackage{ulem}
\usepackage{longtable}
\usepackage{microtype}

\graphicspath{{./}{figures/}}
\newcommand{\CfA}{Center for Astrophysics \textbar{} Harvard \& Smithsonian, 60 Garden Street, Cambridge, MA 02138-1516, USA}

\newcommand{\CIERA}{Center for Interdisciplinary Exploration and Research in Astrophysics and Department of Physics and Astronomy, \\Northwestern University, 2145 Sheridan Road, Evanston, IL 60208-3112, USA}
\newcommand{\WIS}{Department of Particle Physics and Astrophysics, Weizmann Institute of Science, 76100 Rehovot, Israel}
\newcommand{\ESO}{European Southern Observatory, Alonso de C\'ordova 3107, Casilla 19, Santiago, Chile}
\newcommand{\UCSB}{Department of Physics, University of California, Santa Barbara, CA 93106-9530, USA}
\newcommand{\LCO}{Las Cumbres Observatory, 6740 Cortona Drive, Suite 102, Goleta, CA 93117-5575, USA}

\newcommand{\Southampton}{Department of Physics and Astronomy, University of Southampton, Southampton SO17 1BJ, UK}

\newcommand{\Carnegie}{Observatories of the Carnegie Institute for Science, 813 Santa Barbara Street, Pasadena, CA 91101-1232, USA}
\newcommand{\Edinburgh}{Institute for Astronomy, University of Edinburgh, Royal Observatory, Blackford Hill EH9 3HJ, UK}
\newcommand{\Birmingham}{Birmingham Institute for Gravitational Wave Astronomy and School of Physics and Astronomy, University of Birmingham, Birmingham B15 2TT, UK}
\newcommand{\Bath}{Department of Physics, University of Bath, Claverton Down, Bath BA2 7AY, UK}

\newcommand{\Ohio}{Astrophysical Institute, Department of Physics and Astronomy, 251B Clippinger Lab, Ohio University, Athens, OH 45701-2942, USA}

\newcommand{\Granada}{Departamento de F\'isica Te\'orica y del Cosmos, Universidad de Granada, E-18071 Granada, Spain}
\newcommand{\TelAviv}{The School of Physics and Astronomy, Tel Aviv University, Tel Aviv 69978, Israel}
\newcommand{\CIFAR}{CIFAR Azrieli Global Scholars program, CIFAR, Toronto, Canada}

\title[The TDE AT\,2018hyz]{The Tidal Disruption Event AT\,2018hyz II: Light Curve Modeling of a Partially Disrupted Star}

\author[S. Gomez et al]{\href{https://orcid.org/0000-0001-6395-6702}{Sebastian Gomez$^{1}$}\thanks{Contact e-mail: \href{mailto:sgomez@cfa.harvard.edu}{sgomez@cfa.harvard.edu}}, 
\href{https://orcid.org/0000-0002-2555-3192}{Matt Nicholl$^{2,3}$},
\href{https://orcid.org/0000-0002-5096-9464}{Philip Short$^{3}$},
\href{https://orcid.org/0000-0003-4768-7586}{Raffaella Margutti$^{4}$},
\newauthor 
\href{https://orcid.org/0000-0003-0526-2248}{Kate D.~Alexander$^{4}$\thanks{Einstein Fellow}},
\href{https://orcid.org/0000-0003-0526-2248}{Peter K.~Blanchard$^{4}$},
\href{https://orcid.org/0000-0002-9392-9681}{Edo Berger$^{1}$},
\href{https://orcid.org/0000-0003-0307-9984}{Tarraneh Eftekhari$^{1}$},
\newauthor 
\href{https://orcid.org/0000-0001-6797-1889}{Steve Schulze$^{5}$},
\href{https://orcid.org/0000-0003-0227-3451}{Joseph Anderson$^{6}$},
\href{https://orcid.org/0000-0001-7090-4898}{Iair Arcavi$^{7,8}$},
\href{https://orcid.org/0000-0002-7706-5668}{Ryan Chornock$^{9}$},
\newauthor 
\href{https://orcid.org/0000-0002-2478-6939}{Philip S.~Cowperthwaite$^{10}$},
\href{https://orcid.org/0000-0002-1296-6887}{Llu\'is Galbany$^{11}$},
\href{https://orcid.org/0000-0002-3680-9712}{Laura J. Herzog$^{9}$},
\newauthor 
\href{https://orcid.org/0000-0002-1125-9187}{Daichi Hiramatsu$^{12,13}$},
\href{https://orcid.org/0000-0002-0832-2974}{Griffin Hosseinzadeh$^{1}$},
\href{https://orcid.org/0000-0003-1792-2338}{Tanmoy Laskar$^{14}$},
\newauthor
\href{https://orcid.org/0000-0003-3939-7167}{Tom\'as E. M\"uller Bravo$^{15}$},
\href{https://orcid.org/0000-0002-7640-236X}{Locke Patton$^{1}$} and
\href{https://orcid.org/0000-0003-0794-5982}{Giacomo Terreran$^{4}$}
\\
$^{1}$\CfA\\
$^{2}$\Birmingham\\
$^{3}$\Edinburgh\\
$^{4}$\CIERA\\
$^{5}$\WIS\\
$^{6}$\ESO\\
$^{7}$\TelAviv\\
$^{8}$\CIFAR\\
$^{9}$\Ohio\\
$^{10}$\Carnegie\\
$^{11}$\Granada\\
$^{12}$\LCO\\
$^{13}$\UCSB\\
$^{14}$\Bath\\
$^{15}$\Southampton
}

\date{Accepted 2020 July 10. Received 2020 July 10; in original form 2020 March 11}

\pubyear{2020}

\begin{document}
\label{firstpage}
\pagerange{\pageref{firstpage}--\pageref{lastpage}}
\maketitle
\vspace{-1.5cm}

\begin{abstract}
AT\,2018hyz (=ASASSN-18zj) is a tidal disruption event (TDE) located in the nucleus of a quiescent E+A galaxy at a redshift of $z = 0.04573$, first detected by the All-Sky Automated Survey for Supernovae (ASAS-SN). We present optical+UV photometry of the transient, as well as an X-ray spectrum and radio upper limits. The bolometric light curve of AT\,2018hyz is comparable to other known TDEs and declines at a rate consistent with a $t^{-5/3}$ at early times, emitting a total radiated energy of $E = 9\times10^{50}$ erg. An excess bump appears in the UV light curve about 50 days after bolometric peak, followed by a flattening beyond 250 days. We detect a constant X-ray source present for at least 86 days. The X-ray spectrum shows a total unabsorbed flux of \mbox{$\sim 4\times10^{-14}$ erg cm$^{-2}$ s$^{-1}$} and is best fit by a blackbody plus power-law model with a photon index of $\Gamma = 0.8$. A thermal X-ray model is unable to account for photons $> 1$ keV, while a radio non-detection favors inverse-Compton scattering rather than a jet for the non-thermal component. We model the optical and UV light curves using the Modular Open-Source Fitter for Transients ({\tt MOSFiT}) and find a best fit for a black hole of $5.2\times10^6$ M$_\odot$ disrupting a $0.1$ M$_\odot$ star; the model suggests the star was likely only partially disrupted, based on the derived impact parameter of $\beta=0.6$. The low optical depth implied by the small debris mass may explain how we are able to see hydrogen emission with disk-like line profiles in the spectra of AT\,2018hyz (see our companion paper, \citealt{Short20}).\vspace{-0.2cm}
\end{abstract}
\begin{keywords}
galaxies: nuclei -- black hole physics -- transients:  tidal disruption events\vspace{-0.4cm}
\end{keywords}

\section{Introduction} \label{sec:intro}
\vspace{-0.3cm}
A tidal disruption event (TDE) can occur when a star gets too close to a supermassive black hole such that the tidal forces from the black hole exceed the self-gravity of the star, eventually tearing it apart \citep{hills75,Rees88}. Following this disruption, the material from the star is expected to circularize into an accretion disk, and a fallback accretion phase begins, powering an optical transient \citep{Gezari09, Guillochon09}. There are about 60 known TDEs, showing a wide gamut of observational features \citep{Auchettl17, Mockler19, Velzen20}. Some exhibit hydrogen and helium emission, while others only helium \citep{Gezari12, Arcavi14}. More recently, TDEs with nitrogen and oxygen lines, powered by Bowen fluorescence, have been detected \citep{Blagorodnova19, Leloudas19}. \cite{Velzen20} defined three classes: TDE-H (hydrogen only), TDE-He (helium only) and TDE-Bowen (Bowen lines in combination with H and/or He). At least one TDE has evolved from showing hydrogen and Bowen lines to helium-only \citep{Nicholl19_17eqx}. Some TDEs show X-ray emission in excess of the optical luminosity, while others are X-ray dim \citep{Holoien16,Auchettl17}. Additionally, radio observations suggest a few TDEs drive relativistic outflows, while others do not \citep{Zauderer11, Bower13, Velzen13, Alexander16}.

In this paper we present radio, optical, UV, and X-ray observations of AT\,2018hyz, originally discovered as a nuclear optical transient by the All-Sky Automated Survey for Supernovae (\mbox{ASAS-SN}; \citealt{Shappee14, Kochanek}) on 2018 Nov 6 and designated ASASSN-18zj \citep{Brimacombe18}. The transient was classified as a TDE on 2018 Nov 9 by \cite{Dong18} and independently verified by \cite{Arcavi18} on 2018 Nov 12. \cite{Velzen20} first presented optical+\textit{Swift} photometry of AT\,2018hyz. The authors classify it as a TDE-H, one with broad H$\alpha$ and H$\beta$ lines. The early hydrogen-dominated spectrum transitions to being helium dominated \citep{Short20}. The spectra are also blue and show broad double-peaked emission lines that evolve in shape, from a smooth broad profile, to boxy, and then smooth again. For an in-depth description of the spectra see \cite{Short20}.

In \S\ref{sec:host} we describe the host galaxy of AT\,2018hyz. In \S\ref{sec:observations} we present our follow-up observations and describe the publicly available observations of AT\,2018hyz. In \S\ref{sec:modeling} we present our modeling of the light curve. In \S\ref{sec:properties} we outline different properties of the light curves, and in \S\ref{sec:conclusion} we outline our key conclusions. Throughout this paper we assume a flat $\Lambda$CDM cosmology with \mbox{$H_{0} = 69.3$ km s$^{-1}$ Mpc$^{-1}$}, $\Omega_{m} = 0.286$, and $\Omega_{\Lambda} = 0.712$ \citep{Hinshaw13}.

\section{Host Galaxy} \label{sec:host} 

From an archival SDSS spectrum, we see that AT\,2018hyz is found in the nucleus of a quiescent E+A galaxy \citep{Short20}. It is unsurprising to see a TDE in this galaxy, since it has been shown that TDEs tend to be over-represented in these types of galaxies by a factor of $30-35$ \citep{Arcavi14,French16,Graur18}. \cite{Velzen20} modeled the host galaxy photometry (Table~\ref{tab:host}) with {\tt Prospector} \citep{Leja17} and find a host mass of $\log{(M / M_\odot)} = 9.84^{+0.09}_{-0.14}$, a stellar population age of $4.74^{+2.98}_{-1.40}$ Gyr and a metallicity of $Z/Z_\odot = -1.41^{+0.44}_{-0.37}$.

\begin{figure}
	\begin{center}
		\includegraphics[width=1.0\columnwidth]{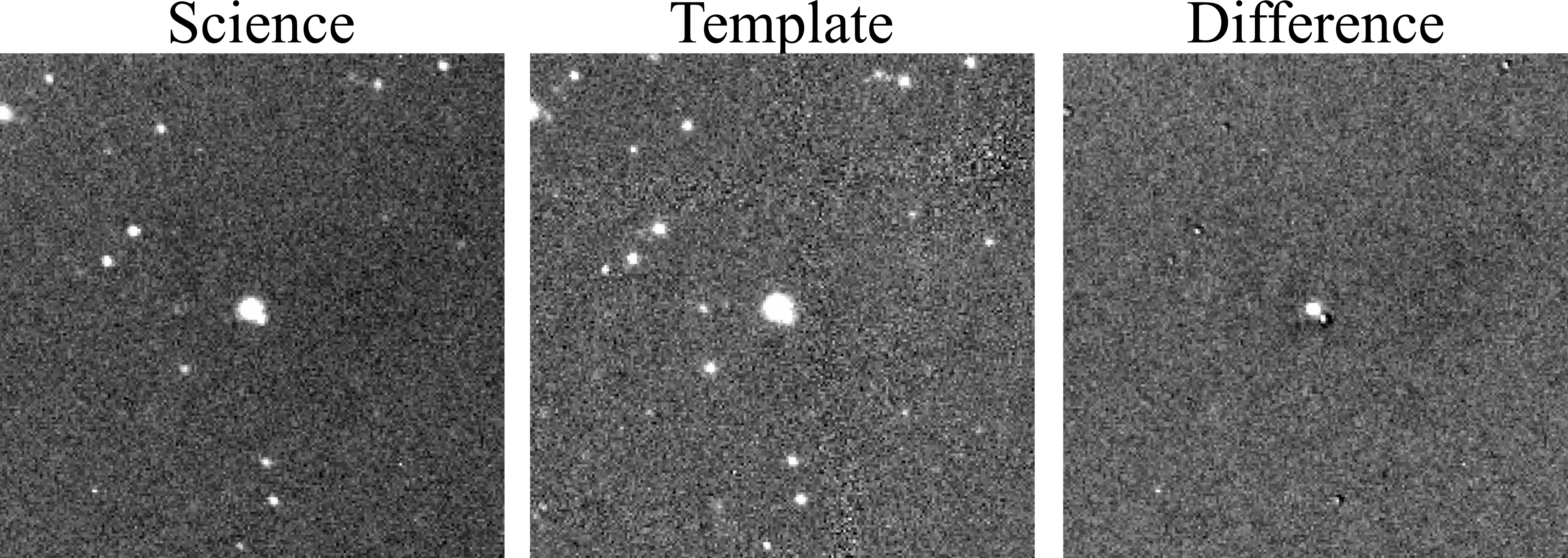}
		\caption{Image of AT\,2018hyz taken with KeplerCam in i-band ({\it Left}), an archival template image from PS1/$3\pi$ ({\it Middle}), and the difference of the two images, where the transient is clearly visible ({\it Right}). \label{fig:difference}}	
	\end{center}
\end{figure}

We model the host's SED in order to derive its magnitude in the UVOT bands. We performed photometry on images of the host from SDSS and $3\pi$, using a $5''$ aperture for all filters. We generated $3.9\times10^6$ templates based on the \citet{Bruzual2003a} stellar population-synthesis models with the Chabrier initial mass function \citep[IMF;][]{Chabrier2003a}. The star formation history (SFH) was approximated by a declining exponential function of the form $\exp \left(-t/\tau\right)$, where $t$ is the age of the stellar population and $\tau$ the e-folding time-scale of the SFH (varied in nine steps between 0.1 and 30 Gyr). These templates were attenuated with the \citet{Calzetti2000a} model that we varied in 22 steps from $E(B-V)=0$ to 1 mag. The best-fitting templates were identified from $\chi^2$ minimization. Excluding NIR photometry (due to contamination by a nearby red star), we find a mass $\log{(M / M_\odot)} = 9.40^{+0.56}_{-0.12}$, in broad agreement with \citet{Velzen20}, and negligible current star formation. The limits on star formation rate of \mbox{$<8.6\times10^{-6}$ M$_\odot$ yr$^{-1}$} placed by our ALMA non-detections rule out obscured star-formation, in agreement with the current classification as an E+A galaxy and the lack of star-formation inferred from our model fit. We convolve the SED of the best-fitting templates with the UVOT passbands to derive the estimated magnitude of the host in these bands, shown in Table~\ref{tab:host}.

\section{Observations} \label{sec:observations}

\subsection{Optical and UV Photometry}\label{sec:optical} 

\begin{table}
\caption{List of publicly available photometry of the host galaxy. The magnitudes are corrected for Galactic extinction. The UVOT model magnitudes are derived from the host's SED model described in \S\ref{sec:host}.}
\label{tab:host}
\begin{tabular}{ccc}
\hline
  & Value & Units\\
\hline
    NUV       & $21.57  \pm 0.26  $  & mag ( GALEX) \\
    $u$       & $19.06  \pm 0.04  $  & mag ( SDSS ) \\
    $g$       & $17.49  \pm 0.01  $  & mag ( SDSS ) \\
    $g$       & $17.46  \pm 0.01  $  & mag ( 3PI  ) \\
    $r$       & $16.96  \pm 0.01  $  & mag ( SDSS ) \\
    $r$       & $16.98  \pm 0.02  $  & mag ( 3PI  ) \\
    $i$       & $16.69  \pm 0.01  $  & mag ( SDSS ) \\
    $i$       & $16.71  \pm 0.02  $  & mag ( 3PI  ) \\
    $z$       & $16.55  \pm 0.01  $  & mag ( 3PI  ) \\
    $z$       & $16.51  \pm 0.01  $  & mag ( SDSS ) \\
    $y$       & $16.44  \pm 0.01  $  & mag ( 3PI  ) \\
    V         & 17.09                & model mag    \\
	B         & 17.74                & model mag    \\
	U         & 19.12                & model mag    \\
	UVW1      & 20.73                & model mag    \\
	UVM2      & 21.31                & model mag    \\
	UVW2      & 21.76                & model mag    \\
\hline
\end{tabular}
\end{table}

AT\,2018hyz was first detected by ASAS-SN on 2018 Oct 14 with a magnitude of $g = 17.08 \pm 0.22$ and a prior non-detection of $g > 16.16$ on 2018 Oct 10, with no previous deeper upper limits \citep{Shappee14, Kochanek}. ASAS-SN observed AT\,2018hyz regularly until 2019 Jun 27 and provided $g$ and V band measurements of the source. The ASAS-SN photometry used in this work was obtained from the ASAS-SN Sky Patrol database \footnote{\url{https://asas-sn.osu.edu/}}. We average the ASAS-SN photometry on bins of 1 day and only make use of the data before and during peak (extending to \mbox{${\rm MJD} = 58446$}) due to a large observed scatter in the later data.

\begin{figure}
	\begin{center}
		\includegraphics[width=1.0\columnwidth]{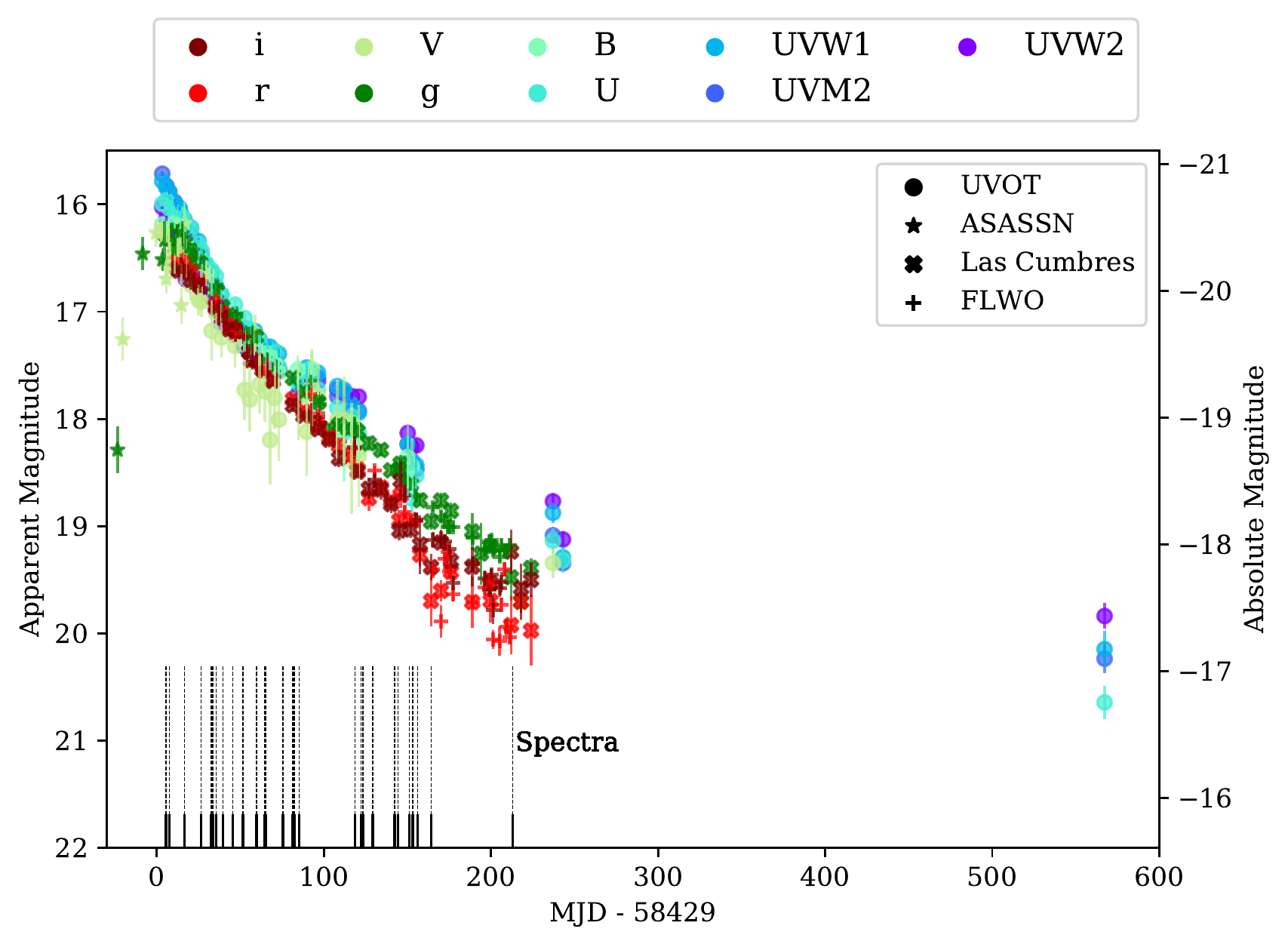}
		\caption{Optical and UV light curves of AT\,2018hyz, host-subtracted and corrected for galactic extinction. The black lines mark the times for which we have optical spectra \citep{Short20}. The photometry shown here is available on the online version of this journal. \label{fig:lightcurve}}	
	\end{center}
\end{figure}

The UV-Optical Telescope (UVOT; \citealt{Roming05}) on board the Neil Gehrels \textit{Swift} observatory (\textit{Swift}; \citealt{Gehrels04}) regularly observed AT\,2018hyz from 2018 Nov 10 until 2019 March 07 with a cadence of about 3 days, with further observations extending to 2019 July 08 until the source became sun-constrained \mbox{(ObsID: 000109750[01-38])}. We requested an additional late-time epoch and observed the source on 2020 May 22, 28 and 31. We determined the magnitude of AT\,2018hyz by performing aperture photometry with the HEAsoft {\tt uvotsource} function \citep{Nasa14}. We use a $5''$ aperture centered on the target to extract UVW2, UVM2, UVW1, U, V, and B transient+host magnitudes and a $25''$ region to determine background statistics.

In order to isolate the magnitude of the transient, we subtract the magnitude of the host galaxy from each UVOT and ASAS-SN measurement in the corresponding filter, calculated in \S\ref{sec:host}. All the photometry reported in this work is calibrated to AB magnitudes.

We obtained images of AT\,2018hyz in $gri$ filters using KeplerCam on the 1.2-m telescope at Fred Lawrence Whipple Observatory (FLWO) and the Las Cumbres Observatory’s network of 1m telescopes \citep{Brown13}. We processed the images using standard {\tt IRAF}\footnote{\label{IRAF}{\tt IRAF} is written and supported by the National Optical Astronomy Observatories, operated by the Association of Universities for Research in Astronomy, Inc. under cooperative agreement with the National Science Foundation.} routines, and performed photometry with the {\tt daophot} package. Instrumental magnitudes were measured by modeling the point-spread function (PSF) of each image using reference stars in the image. For calibration, we estimated individual zero-points of each image by measuring the magnitudes of field stars and comparing to photometric AB magnitudes from the PS1/$3\pi$ catalog. The uncertainties reported in this work are the combination of the photometric uncertainty and the zero-point determination uncertainty. To isolate AT\,2018hyz from its host galaxy we perform image subtraction on each $gri$ image using {\tt HOTPANTS} \citep{Becker15}. We use archival PS1/$3\pi$ images as reference templates \citep{Chambers18}; an example is shown in Figure~\ref{fig:difference}.

We note there is a red star $\sim 5\farcs$ away from the center of the host galaxy. After performing image subtraction and PSF photometry on the resulting image, we do not detect contamination from the star in the optical bands. Given that this star is brightest in $z$ band, and $> 2.5$mag fainter in $u$ band, we assume minimal to null contamination in the \textit{Swift} bands. This star does however pose a major contamination problem in infrared photometry, which we do not include in our analysis.

All the ASAS-SN, UVOT, FLWO and Las Cumbres data were corrected for Milky Way galactic extinction using \mbox{$A_V = 0.0917$ mag}, determined using the \cite{Schlafly11} dust maps. The photometry was then corrected to the rest frame from $z = 0.04573$ and shifted in time to define phase 0 as the date of peak bolometric brightness, \mbox{${\rm MJD} = 58429$}. All the optical+UV photometry used for this work is shown in Figure~\ref{fig:lightcurve} and \ref{fig:flux}. The individual FLWO, Las Cumbres, and UVOT data are available in machine readable format in the online version of this journal and on the Open TDE Catalog \footnote{\url{https://tde.space/}} \citep{Guillochon17}.

\subsection{Astrometry}\label{sec:astrometry}

AT\,2018hyz is located in the nucleus of 2MASS J10065085+0141342, a galaxy at a redshift of $z = 0.04573$ or a luminosity distance of $205$ Mpc. We performed astrometry on an FLWO $i$-band image by cross-matching the positions of field stars in the image to the ICRS coordinates from \mbox{\textit{Gaia}-DR2} \citep{Gaia16, Gaia18}. We measure the centroid of AT\,2018hyz on template subtracted images to be R.A.=${\rm 10^h06^m50^s.872}$, decl.=$+01^\circ41'34\farcs10$ (J2000), with a centroid uncertainty of $0\farcs24$. We perform relative astrometry to measure the separation between AT\,2018hyz and the center of its host galaxy. Using a pre-explosion template from archival PS1/$3\pi$ images as reference, and 10 template subtracted FLWO $g$-band images to measure the position of AT\,2018hyz. The resulting offset is $0\farcs2 \pm 0\farcs8$, equivalent to a physical separation of $0.2\pm 0.8$ kpc, consistent with the transient being nuclear. Where the uncertainty is the combination of the astrometric error and the scatter in the measured position among different images.

\subsection{Radio and Millimeter Observations}\label{sec:radio} 

We obtained millimeter observations with the Atacama Large Millimeter/submillimeter Array (ALMA) in Band 3 ($\sim 100$ GHz) on 2018 November 28 and December 19 with a total on-source integration time of 22.2 minutes per epoch. We report the results of the ALMA data products which used J1058+0133 for bandpass and flux density calibration and J1010-0200 for complex gain calibration. The November 28 and December 19 epochs were imaged using 840 and 378 pixels, respectively, with an image scale of 0.13 and 0.29 arcsec per pixel, corresponding to a synthesized beam size of $0.81'' \times 0.68''$ and $1.98'' \times 1.47''$, respectively. The images were created using multi-frequency synthesis (MFS; \citealt{Sault94}), Briggs weighting with a robust parameter of 0.5, and a standard gridding convolution function. The source is not detected in either epoch with a $3\sigma$ limit of $F_\nu (\rm 100 GHz)$ $\lesssim 37.6$ and $42.7 \mu$Jy for the November and December observations, respectively. This corresponds a to very low star formation rate of \mbox{$<8.6\times10^{-6}$ M$_\odot$ yr$^{-1}$} \citep{Kennicutt98}.

\begin{figure}
	\begin{center}
		\includegraphics[width=\columnwidth]{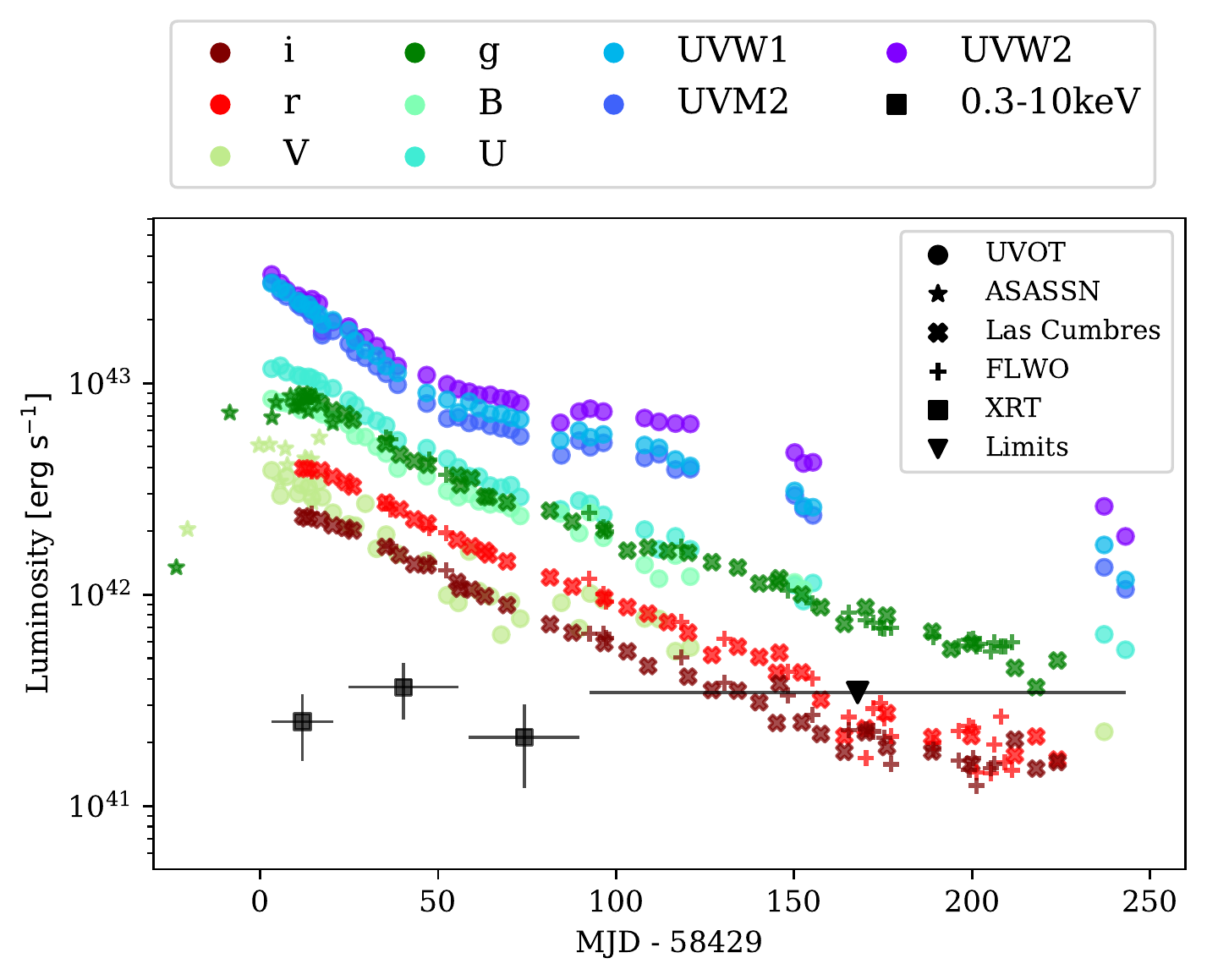}
		\caption{Optical and UV light curve of AT\,2018hyz, excluding the epoch at 550d for clarity. The points in black are three binned \textit{Swift}-XRT observations showing the unabsorbed flux and one upper limit. \label{fig:flux}}
	\end{center}
\end{figure}

Four hours of AMI-LA 15.5 GHz observations revealed no radio source at the location of the transient, corresponding to a 3$\sigma$ upper limit of \mbox{$\sim 85$ $\mu$Jy} on 2018 Nov 15 \citep{Horesh18}. This upper limit corresponds to a luminosity $\nu L_\nu<6.6\times10^{37}$ erg s$^{-1}$, slightly deeper than the radio detections for ASASSN-14li, possibly a jetted TDE \citep{Velzen16,Alexander16}. Our limits are comparable to some of the deepest radio upper limits for other TDEs, such as AT\,2018zr \citep{Velzen19_2018zr} and AT\,2017eqx \citep{Nicholl19_17eqx}, though shallower than iPTF16fnl \citep{Blagorodnova17}; these had no detected radio counterparts. 

\mbox{ASASSN-14li} has a ratio of total thermal energy to jet energy of $10^{2.5}$ \citep{Velzen16}. If AT\,2018hyz has a similar ratio, the total optical+UV energy of \mbox{$\sim 10^{51}$} erg in AT\,2018hyz would imply a jet energy of \mbox{$\sim 10^{48}$ erg}. Our radio non-detection of AT\,2018hyz suggests that any outflow may have been less energetic than that in \mbox{ASASSN-14li}; however, a lower ambient density or larger off-axis observing angle could also be responsible for the lower radio luminosity in this event.

\subsection{X-ray Observations}\label{sec:x-ray} 

AT\,2018hyz was observed by the X-ray Telescope (XRT) onboard \textit{Swift} \citep{Burrows05}. We reduced the \textit{Swift}-XRT data following the prescriptions by \cite{Margutti13} with HEAsoft v6.26.1 and corresponding calibration files. We apply standard filtering criteria and bin the data into four separate epochs to estimate the source count-rate evolution with time. An X-ray source is detected in three epochs up to a phase of 86 days, while the source is not detected after binning all the data from a phase of 86 to 232 days. The upper limit obtained from the last bin is shallower that the previous detection and is therefore unconstraining, the flux-calibrated X-ray light curve is shown in Figure~\ref{fig:flux}. An additional 3ks observation obtained in 2020 May (phase $\sim 550$ days) yielded a non-detection with an unconstraining upper limit of $4.4\times10^{-13}$ erg cm$^{-2}$ s$^{-1}$.

\begin{figure}
	\begin{center}
		\includegraphics[width=1.0\columnwidth]{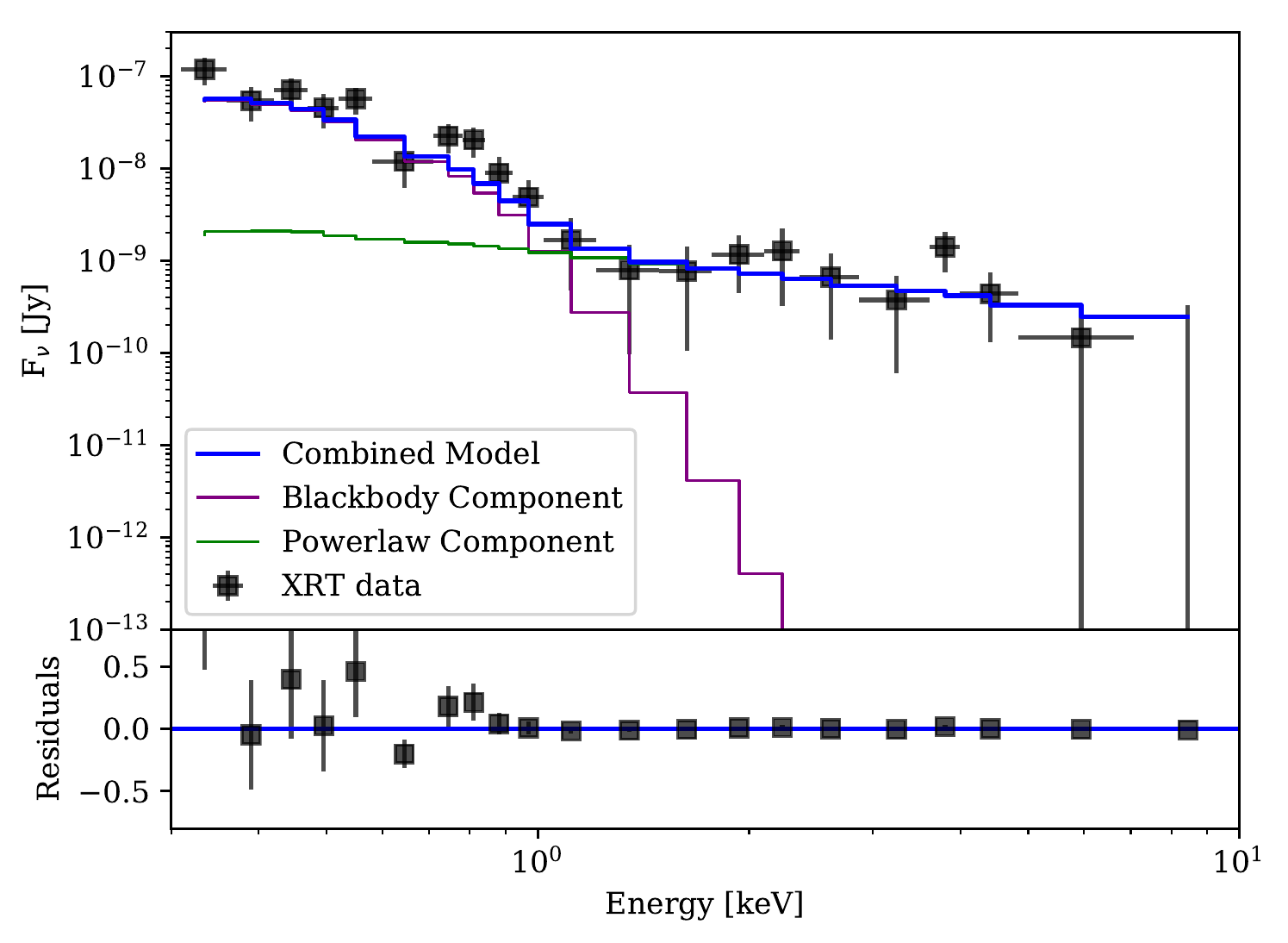}
		\caption{Combined \textit{Swift}-XRT spectra of AT\,2018hyz extending to a phase of 86 days and fitted with an absorbed blackbody plus power-law model. A blackbody spectrum is unable to account for the high energy photons and is therefore disfavored. The blue line is the sum of the individual components, shown in purple and green. \label{fig:xray}}
	\end{center}
\end{figure}

\begin{figure}
	\begin{center}
		\includegraphics[width=\columnwidth]{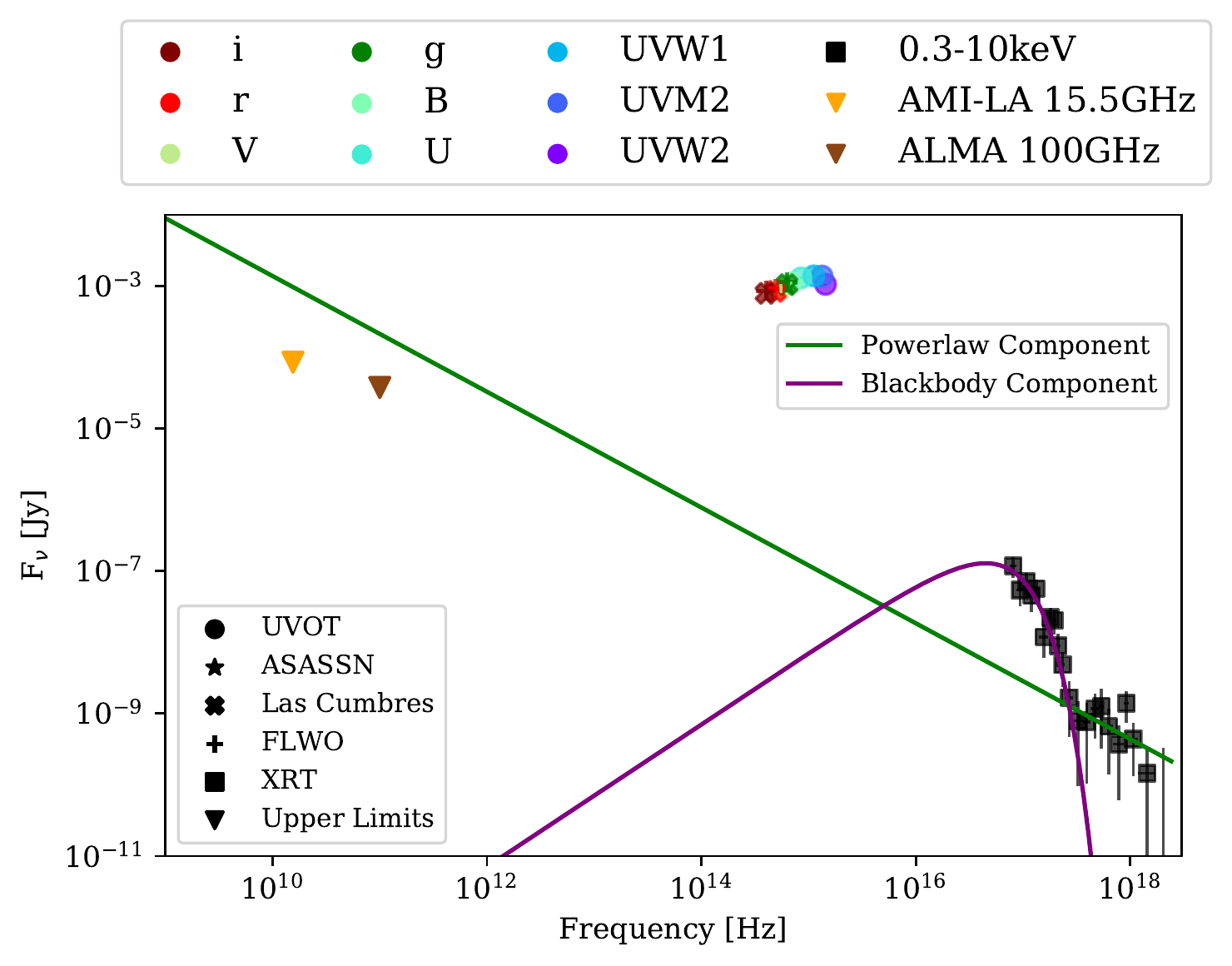}
		\caption{Broadband SED of AT\,2018hyz, using the optical and UV photometry near peak, between a phase of 0 and 25 days. Radio upper limits are from our ALMA observations and one AMI-LA observation from \citet{Horesh18}. The green line shows an extrapolation of the powerlaw component of the X-ray model to the radio, and the purple line shows the blackbody component of the X-ray model.
		\label{fig:powerlaw}}
	\end{center}
\end{figure}

We do not find evidence for a statistically significant spectral evolution of the source. We extract a spectrum comprising the data at $t<86$ days to constrain the spectral properties of the source and the count-to-flux conversion factor. We fit the $0.3-10$ keV spectrum with a single absorbed power-law model with {\tt XSPEC} and find a best fit photon index of $\Gamma = 3.2 \pm 0.3$ with no evidence for intrinsic absorption. We also fit the X-ray spectrum with a single blackbody and find a best fit temperature of T$ = 0.12$ keV, this model is unable to account for high energy photons above $>1$ keV, and is therefore disfavored. In Figure~\ref{fig:xray} we show our preferred model, where we fit the spectrum with a blackbody component and an additional power law to account for the high energy photons. For this model we find a best fit to the blackbody component of T$ = 0.11 \pm 0.03$ keV, and a photon index of $\Gamma = 0.8\pm 0.6$. We adopt a redshift of $z = 0.04573$ and a neutral hydrogen column density along the line of sight to AT\,2018hyz of NH$_{mw} = 2.67\times10^{20}$ cm$^{-2}$ \citep{Kalberla05}. The total unabsorbed flux is $4.1^{+0.6}_{-0.4}\times10^{-14}$ erg cm$^{-2}$ s$^{-1}$ ($1\sigma$ errors). For this spectrum, the count-to-flux conversion factor is $6.16\times10^{-11}$ erg cm$^{-2}$ counts$^{-1}$.

Most TDEs show a decline in their X-ray light curve \citep{Auchettl17}. Other TDEs, such as ASASSN-15oi, show a flat X-ray light curve with a subsequent late-time X-ray brightening. Early observations of ASASSN-15oi by \cite{Holoien16_15oi} can not rule out a low-luminosity AGN; while late observations of the same source from \cite{Gezari17} show brightening consistent with the X-ray emission from a thermal accretion disk from the TDE. AT\,2018fyk is another example of a TDE with a flat X-ray light curve with a subsequent brightening \citep{Wevers19}. The X-ray luminosity of AT\,2018hyz is not declining, but consistent with being flat. The X-rays in AT\,2018hyz are also consistent with an AGN \citep{Aird15}; given that the measured luminosity of $\approx 3\times 10^{41}$\,erg\,s$^{-1}$ is $\sim0.1-1\%$ of the Eddington luminosity for the inferred black hole mass of $\sim 10^{6-7}$ M$_\odot$ (See \S\ref{sec:modeling} and \cite{Short20} for a description of the mass estimates). 

The ratio between the [O III] and hard (2-20 keV) X-ray luminosity functions of AGNs is $2.15 \pm 0.51$ dex \citep{Heckman05}. The host of AT\,2018hyz has an archival spectrum from the Sloan Digital Sky Survey (SDSS), with emission line fluxes from the Portsmouth emission line Value Added Catalog \citep{Thomas13}. Using the power law component from our model for the X-rays with $\Gamma = 0.8$, and the host galaxy [O III] flux of $1.2\pm0.4\times10^{-16}$\,erg\,cm$^{-2}$\,s$^{-1}$, we find $\log(L_X/L_{\rm [O III]}) = 2.2 \pm 0.2$. This confirms the X-ray luminosity is consistent with an AGN, and that future temporal variability will be required to determine whether it is of AGN or TDE origin. 

We compare the power-law index and luminosity of AT\,2018hyz to the sample of sources from \citet{Auchettl17}, and see that AT\,2018hyz is similar to other confirmed or likely X-ray TDEs. \citet{Auchettl17} suggest that TDEs can separate into thermal TDEs without a jet, and non-thermal TDEs with a jet. Given the fact that AT\,2018hyz has a non-thermal spectrum, that would be indicative of the presence of a jet, or inverse-Compton scattering of X-ray photons from the accretion disk. We fail to detect a jet in radio observations, but future temporal evolution will distinguish whether the X-ray emission is indeed dominated by the TDE or if it is a weak pre-existing AGN.

In Figure~\ref{fig:powerlaw} we show a broadband SED with the X-ray spectra and best fit model compared to the optical photometry and radio upper limits. We do not account for self-absorption in our extrapolation, however this is reasonable over this frequency range. ASASSN-14li showed a self-absorbed synchrotron spectrum with a peak that moved from $\sim 20$\,GHz to $\sim 2$\,GHz \citep{Alexander16}, i.e.~the turnover was at all times below the frequencies of our radio upper limits. The SED slope measured in radio observations of ASASSN-14li was $F_\nu \propto \nu^{-1}$ (compared to $F_\nu \propto \nu^{-0.2 \pm 0.6}$ for the X-ray model fit to AT\,2018hyz). Using a steeper power-law more similar to ASASSN-14li would give a larger discrepancy between the model prediction and our radio upper limits. Thus the radio limits appear to favour inverse-Compton scattering for the non-thermal X-rays.

\begin{figure}
	\begin{center}
		\includegraphics[width=\columnwidth]{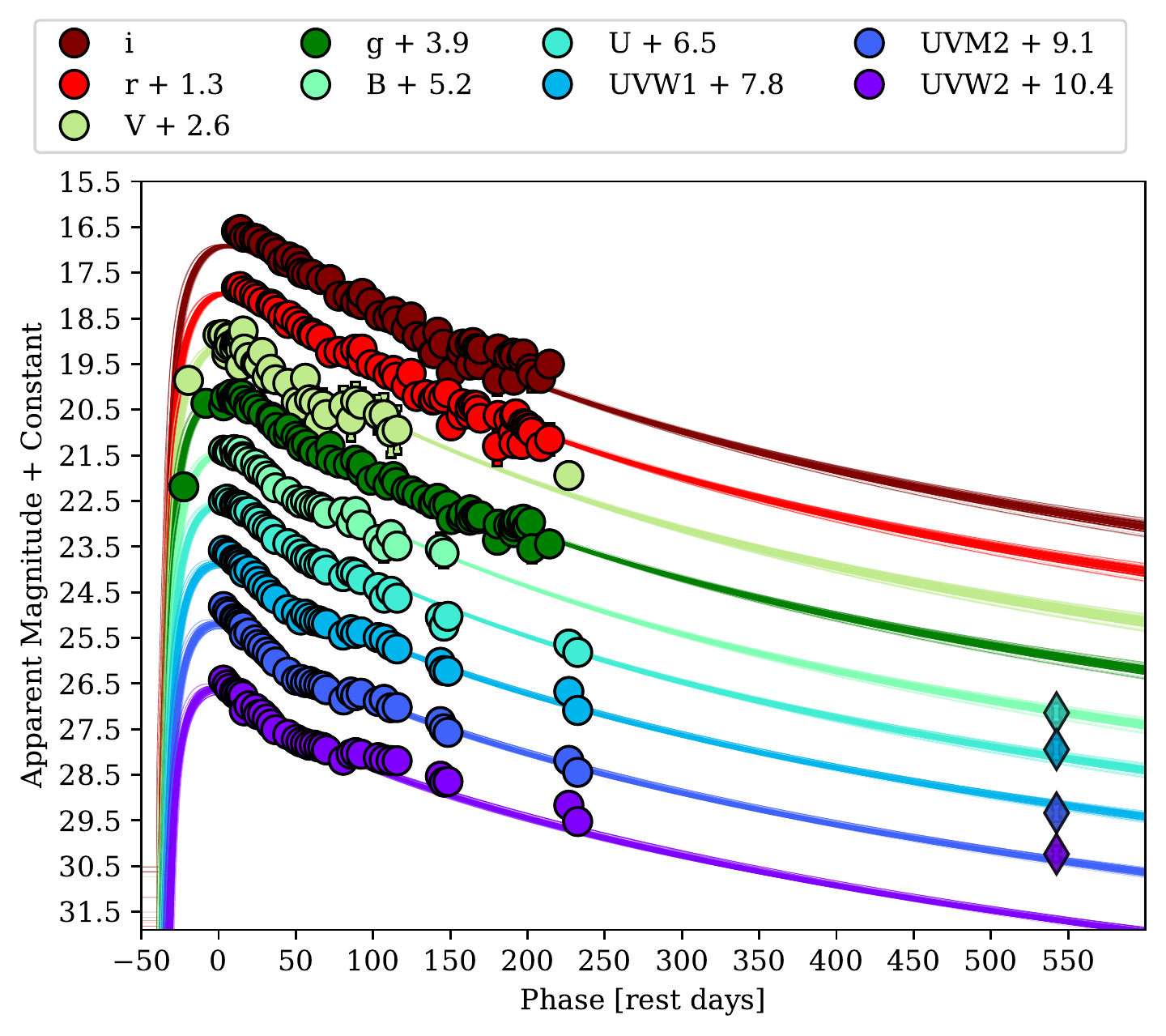}
		\caption{Light curves of AT\,2018hyz with the best model realizations from {\tt MOSFiT}. We fit the light curves with a {\tt TDE} model and list the best fit parameters in Table~\ref{tab:mosfit}. The shaded diamonds at late times were excluded from the fit, since they do not follow the expected fallback rate, but are instead likely due to late-time accretion. \label{fig:mosfit}}
	\end{center}
\end{figure}

\section{Light Curve Modeling} \label{sec:modeling} 

We model the light curves of AT\,2018hyz using the {\tt TDE} model in the {\tt MOSFiT} Python package, a Markov chain Monte Carlo (MCMC) code designed to model the light curves of transients using a variety of different power sources \citep{guillochon18}. The {\tt TDE} model in {\tt MOSFiT} estimates the output luminosity by converting the input fallback rate of material from the disrupted star into radiation via an efficiency parameter. The model also takes into account a normalization and power-law exponent for the photosphere. An impact parameter determines whether the star was partially or entirely disrupted. And a viscous timescale defines how fast the accretion disk forms around the black hole. Lastly, to estimate the magnitude of the transient in each observed band, {\tt MOSFiT} assumes a blackbody SED that is convolved with the passband of each filter. The full details of the {\tt TDE} model are described in \cite{Mockler19}. It should be noted that the use of {\tt MOSFiT} is motivated by speed considerations, which allows us to explore a wide parameter space, but requires the use of simple one-zone models that are not overly complex.

\begin{figure}
	\begin{center}
		\includegraphics[width=\columnwidth]{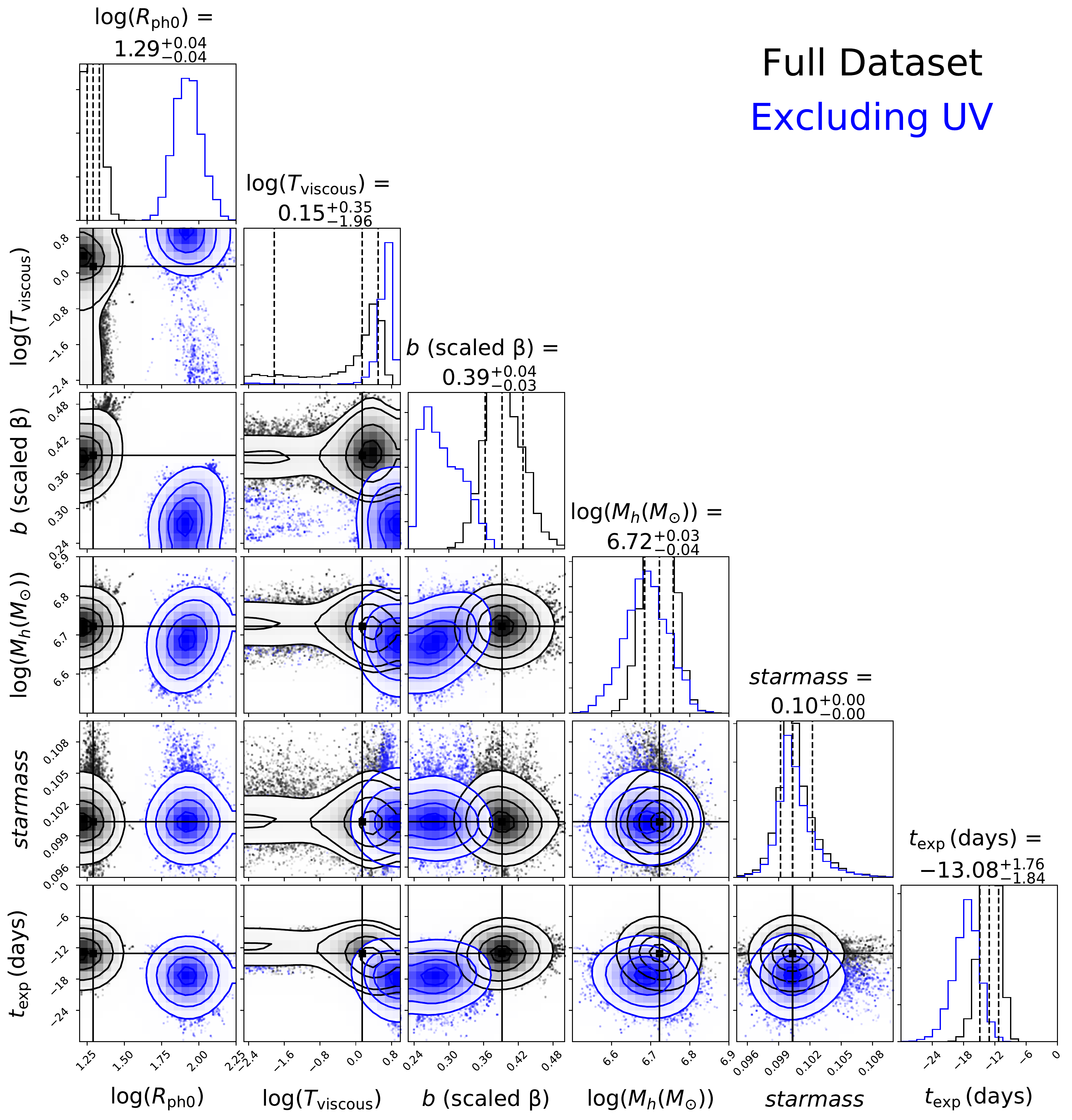}
		\caption{Sample results of the MCMC run for the likely {\tt MOSFiT} models to the light curve of AT\,2018hyz, shown in Figure~\ref{fig:mosfit}. In blue we show the corresponding posteriors when we exclude the $UVW1$, $UVM2$, and $UVW2$ bands. We show the two-dimensional correlation matrices for the parameters being fit. The diagonal shows the marginalized posterior distribution of each parameter. The vertical lines show the highest likelihood estimate for each parameter and the $1\sigma$ error bars. This figure was generated with the {\tt corner} Python package \citep{Foreman16}. \label{fig:corner}}
	\end{center}
\end{figure}

We run the MCMC using an implementation of the {\tt emcee} sampler \citep{foreman13} and test for convergence by ensuring that the models reach a potential scale reduction factor of $<1.2$ \citep{gelman92}, which corresponds to about 2000 steps with 200 walkers. The best-fit parameters of the {\tt TDE} model with the corresponding statistical $1\sigma$ confidence intervals on the fit are shown in Table~\ref{tab:mosfit}. Figure~\ref{fig:mosfit} shows the best model realizations and Figure~\ref{fig:corner} shows the corresponding correlation among the most relevant parameters.

The uncertainties presented in this work represent only the statistical model uncertainties. \cite{Mockler19} quantify the systematic uncertainties of the {\tt MOSFiT} {\tt TDE} model to be $0.66$ dex for the mass of the star, and $0.2$ dex for the mass of the black hole. These uncertainties come mostly from the uncertainty in the mass-radius relation assumed for the disrupted star. The systematic uncertainties of the other parameters being fit are shown in Table~\ref{tab:mosfit}.

\begin{table*}
\caption{Model parameters, flat prior ranges, and 1$\sigma$ error bars for the realizations shown in Figure~\ref{fig:mosfit}. The systematic error is taken from \citet{Mockler19}, determined from assuming different mass-radius relations for disrupted stars. $M_{\rm BH}$ is the mass of the disrupting black hole, $M_*$ is the mass of the disrupted star, $R_{\rm ph0}$ is the photosphere radius power-law normalization coefficient, $T_v$ is the viscous timescale, $b$ is the scaled impact parameter, $\beta$ is the impact parameter, $l$ is the photosphere power-law exponent, $\epsilon$ is the efficiency, $t_{\text{exp}}$ is the disruption time relative to the first data point, $n_{H,\text{host}}$ is the column density in the host galaxy and $A_{V, \text{host}}$ is the corresponding value in extinction, and $\sigma$ is the model uncertainty required to make $\chi^2_r=1$. \newline $\dagger$ These parameters were not fit for, but were calculated using all the posterior distribution samples of the fitted parameters.}

\label{tab:mosfit}
\begin{tabular}{cccccc}
\hline
Parameter & Prior &   Best Fit   &    Best Fit    & Systematic & Units \\
          &       &   (All data) & (Excluding UV) & Error      &       \\
\hline
$ \log{(M_{\rm BH} )}          $ & $[5, 8]      $ & $ 6.72 \pm 0.04           $ & $ 6.69  \pm 0.06       $ & $\pm 0.20$     & M$_\odot$ \\
$ M_*                          $ & $[0.01, 10]  $ & $ 0.100^{+0.002}_{-0.001} $ & $ 0.100 \pm 0.001      $ & $\pm 0.66$ dex & M$_\odot$ \\
$ \log{(R_{\rm ph0} )}         $ & $[-4, 4]     $ & $ 1.29 \pm 0.04           $ & $ 1.92  \pm 0.09       $ & $\pm 0.40$     &           \\
$ \log{(T_v )}                 $ & $[-3, 5]     $ & $ 0.15^{+0.35}_{-1.96}    $ & $ 0.67^{+0.12}_{-0.19} $ & $\pm 0.10$     & days      \\
$ b                            $ & $[0, 2]      $ & $ 0.39 \pm 0.03           $ & $ 0.28^{+0.04}_{-0.03} $ &                &           \\
$ \beta^{\dagger}              $ &                & $ 0.61^{+0.01}_{-0.03}    $ & $ 0.62 \pm 0.02        $ & $\pm 0.35$     &           \\
$ l                            $ & $[0, 4]      $ & $ 0.92 \pm 0.03           $ & $ 1.03 \pm 0.04        $ & $\pm 0.20$     &           \\
$ \epsilon                     $ & $[0.01, 0.4] $ & $ 0.10 \pm 0.02           $ & $ 0.12 \pm 0.03        $ & $\pm 0.68$ dex &           \\
$ t_{\text{exp}}               $ & $[-50, 0]    $ & $ 13.1 \pm 1.8            $ & $ 17.4 \pm 2.5         $ & $\pm 15.0$     & days      \\
$ \log{(n_{H,\text{host}})}    $ & $[16, 23]    $ & $ 17.6 \pm 1.2            $ & $ 17.6 \pm 1.3         $ &                & cm$^{-2}$ \\
$ \log{\sigma}                 $ & $[-4, 2]     $ & $ -0.74 \pm 0.02          $ & $ -0.81 \pm 0.03       $ &                &           \\ 
$ A_{V, \text{host}}^{\dagger} $ &                & $ < 0.01                  $ & $ < 0.01               $ &                & mag       \\
\hline
\end{tabular}
\end{table*}

From the {\tt MOSFiT} model we derive an estimated disruption date of \mbox{${\rm MJD} = 58392 \pm 2$}. We find the best model is that of a black hole of $5.2\times10^6$ M$_\odot$ partially disrupting a star of \mbox{$0.1$ M$_\odot$}. A black hole mass consistent with the value of \mbox{$3.5^{+1.7}_{-0.9}\times10^6$ M$_\odot$} found by \cite{Hung20}. A star is considered partially disrupted when a core remains post-disruption, and fully disrupted when the mass bound to the black hole is greater than half the total mass of the star. The impact parameter $\beta$ (the tidal radius over the pericenter orbit of the star) is reparametrized in {\tt MOSFiT} in terms of the parameter $b$, due to the fact that this varies with the polytrope index of the star and the fraction of mass bound to the black hole. For stars $< 0.3$ M$_\odot$, {\tt MOSFiT} assumes a polytropic index of $\gamma = 5/3$ \citep{Mockler19}. A value of $b = 1$ represents a full disruption of the star, while $b = 0$ signifies no disruption. For AT\,2018hyz we find a value of $b = 0.4$ ($\beta = 0.61$), making this the least disrupted star in the TDE sample of \cite{Mockler19}, with the next lowest value being $\beta = 0.84$ for TDE1. We caution that the systematic uncertainty for $\beta$ in MOSFiT is $\pm 0.35$, which would allow $b$ to vary anywhere from a minimal disruption to a more significant disruption with $b = 0.75$. A full hydrodynamical simulation might be able to confirm if this is the case.

In order to test the robustness of the model and explore its dependence on the UV-bump we run an identical model that excludes the UV data from the fit. With the exception of the photosphere radius, we find the relevant parameters shown in Figure~\ref{fig:corner} to be in good agreement with the model of the full data set, both shown in Table~\ref{tab:mosfit}.

\cite{Ryu20} simulate a series of stars of different masses being disrupted by a $10^6$ M$_\odot$ black hole, similar to the one in AT\,2018hyz. The authors find that for a $0.15$ M$_\odot$ star (their closest model to our best inferred mass of $0.1$ M$_\odot$), an impact parameter of $\beta = 0.61$ corresponds to $> 90\%$ of the star surviving the disruption. For AT\,2018hyz, this would correspond to a disrupted mass of $\lesssim 0.01$ M$_\odot$. For comparison, PS1-11af, a TDE that resulted from the partial disruption of a star \citep{Chornock14}, has a $\beta = 0.90$ and a minimum stripped mass of $\sim 0.006$ M$_\odot$. \cite{Ryu20} also find that for this model the decline rate at late times is closer to $t^{-8/3}$. We show in Figure~\ref{fig:comparisons} how the corresponding $t^{-8/3}$ fit is roughly consistent with the AT\,2018hyz light curve at late times.

For a black hole mass of $10^{6.7}$ M$_\odot$ disrupting a $0.1$ M$_\odot$ star, the implied pericenter of the encounter is $R_p \sim 3 R_{\rm s}$ (or $9 R_{\rm s}$ for \mbox{M $ = 10^{6}$ M$_\odot$}), given the impact parameter of $\beta = 0.6$. The hydrogen emission lines from the accretion disk imply an orbital radius of material emitting at $R\sim600\,R_{\rm s}$ \citep{Short20}. The models of \cite{Bonnerot20} show that the size of the accretion disk in a TDE is $\sim{\rm few} \times10 R_p$. For AT\,2018hyz, the observed emission lines are consistent with this model, but would point towards a smaller black hole mass. For our best estimate of $10^{6.7}$ M$_\odot$, $R \sim 200 R_p$, while for the lower mass black hole estimate of $10^{6}$ M$_\odot$ the radius of the disk would be closer to $R \sim 70 R_p$.

\cite{Short20} find a supermassive black hole mass in the range of \mbox{$1 - 5 \times 10^{6}$ M$_\odot$}, obtained from assuming different M-$\sigma$ relations, lower than the mass estimate of $10^{6.7}$ M$_\odot$ we obtain from \mbox{{\tt MOSFiT}}; which is most similar to the estimate obtained from the \cite{McConnell13} M$_{\rm Bulge}-$ M$_{\rm BH}$ relation of $\sim 10^{6.2}$ M$_\odot$. The peak bolometric luminosity of AT\,2018hyz is $1.9\times10^{44}$ erg s$^{-1}$, corresponding to $\sim0.3L_{\rm Edd}$ for a $10^{6.7}$ M$_\odot$ black hole; this is a typical Eddington ratio for TDE light curve models (see Table~6 in \citealt{Mockler19}).

\section{Observed Properties of the Light Curves}\label{sec:properties} 

\begin{figure}
	\begin{center}
		\includegraphics[width=\columnwidth]{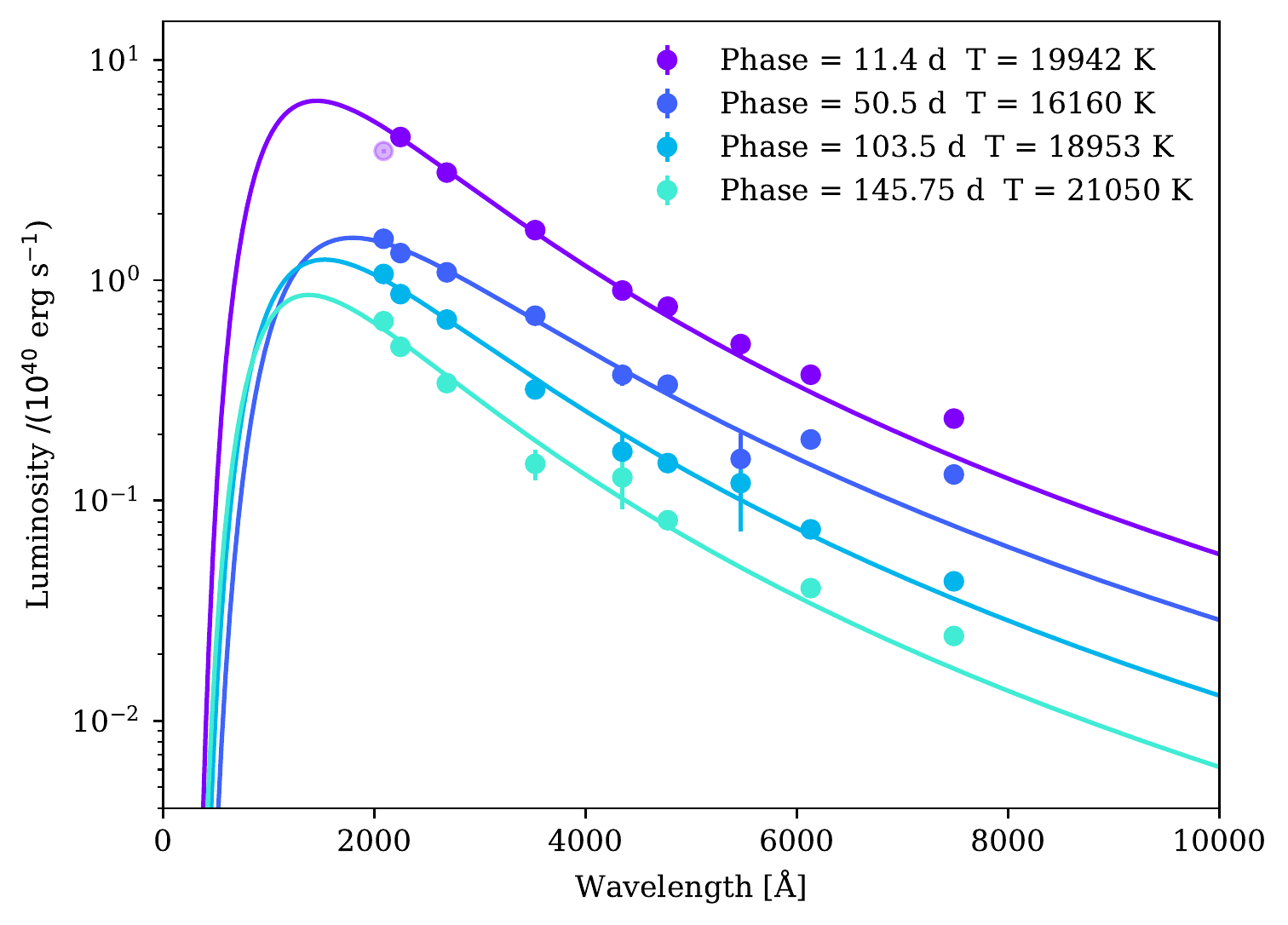}
		\caption{Blackbody fits to four representative epochs of AT\,2018hyz. We see the temperature decline for the first $\sim50$ days, and then rise for $\sim 100$ days. The shaded purple data point is a UVW2 measurement excluded from the fit due to its large deviation from a normal blackbody. This apparent UVW2 suppression might be due to a UV-absorption line, as seen in other TDEs (e.g. \citealt{Blagorodnova19}). \label{fig:seds}}
	\end{center}
\end{figure}

\begin{figure}
	\begin{center}
		\includegraphics[width=\columnwidth]{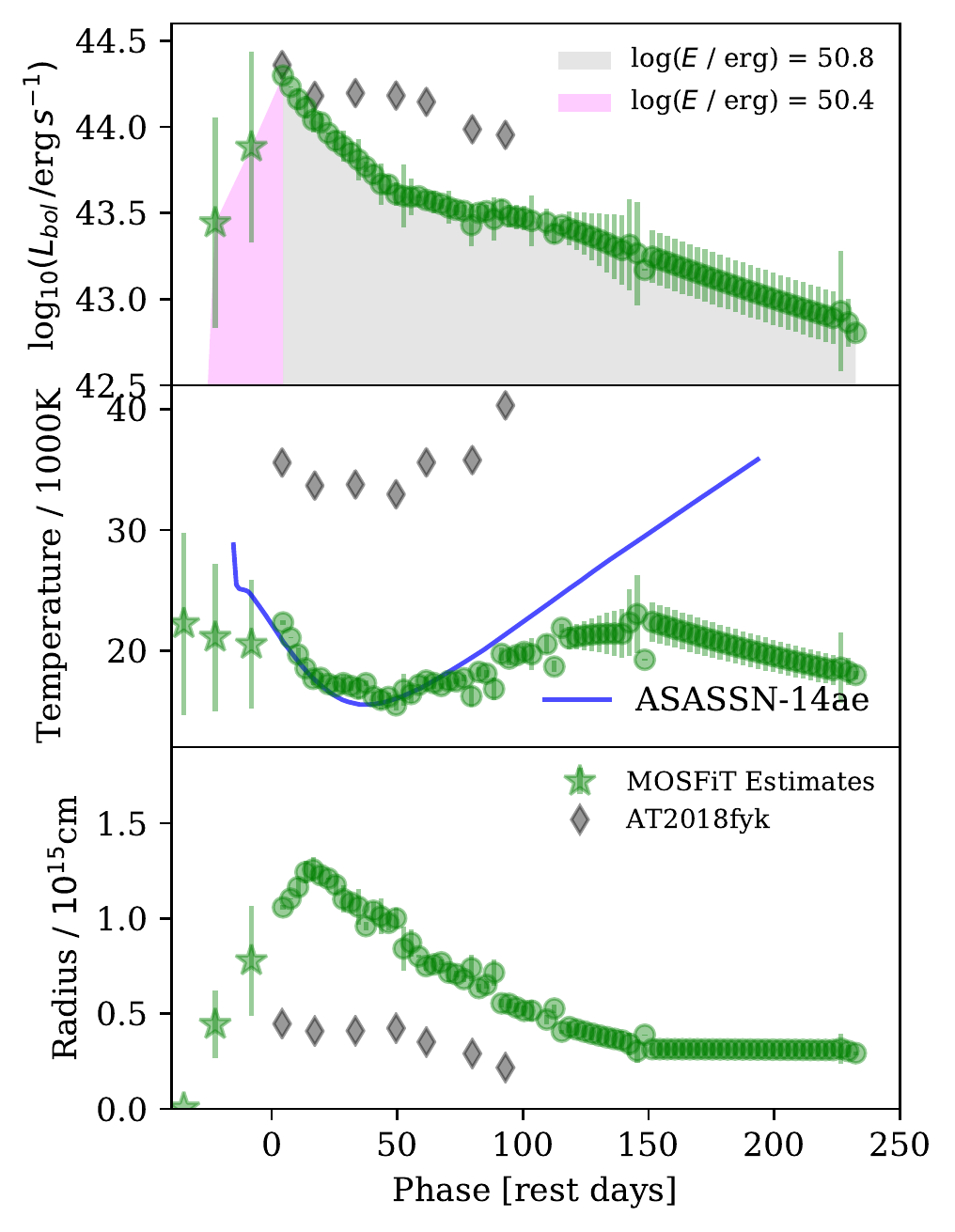}
		\caption{Bolometric light curve ({\it Top}), blackbody temperature ({\it Middle}), and photospheric radius ({\it Bottom}) of AT\,2018hyz. The star-shaped data points are derived from the inferred {\tt MOSFiT} model shown in Figure~\ref{fig:mosfit}. The fits after a phase of 120 days make use of interpolated data to estimate the shape of the SED. The blue line shows the temperature evolution for TDE ASASSN-14ae \citep{Mockler19}, and the gray diamonds are data of the TDE AT\,2018fyk \citep{Wevers19}, shifted in phase to match the luminosity peak of AT\,2018hyz. \label{fig:bolometric}}
	\end{center}
\end{figure}

To obtain the bolometric light curve of AT\,2018hyz we first bin the light curve on three day intervals to be able to generate individual SEDs for each epoch to which we fit a blackbody. For the bins that are missing one or more bands we estimate the missing value by interpolating the full light curves with a 5th or 6th degree polynomial in order to trace the non-monotonic structure of the light curves. We fit a blackbody to each epoch (Figure~\ref{fig:seds}) to measure the bolometric temperature and radius. We estimate the flux outside the observed bands by extrapolating the blackbody fit. We then integrate the entire SED to generate a bolometric light curve. The resulting bolometric light curve, radius and temperature evolution are shown in Figure~\ref{fig:bolometric}.

We calculate the total radiated energy of AT\,2018hyz from a phase of 0 to 233 days to be $E = 6.3\times10^{50}$ erg, obtained by integrating the bolometric light curve shown in Figure~\ref{fig:bolometric}. For the data before a phase of 0 days we lack color information, and therefore estimate the values of luminosity, radius, and temperature from the inferred MOSFiT models described in section~\ref{sec:modeling}. We estimate the total radiated energy before phase of 0 days to be \mbox{$E \approx 2.5\times10^{50}$ erg}. This gives a total radiated energy of $E \approx 9\times10^{50}$ erg for AT\,2018hyz. Similarly, from fitting an empirical model to the light curve, \cite{Velzen20} find a peak luminosity of $\log({\rm L}_g /{\rm erg} / {\rm s}) = 43.57\pm 0.01$, a mean temperature of $\log({\rm T} /{\rm K}) = 4.25\pm 0.01$, and a peak date of MJD$ = 58428$.

We measure a peak temperature of $\sim 22,000$ K near phase 0, which decreases to $\sim 16,000$ K at a phase of 50 days, and then rises back up to $\sim 21,000$ K until a phase of 150 days (Figure~\ref{fig:bolometric}). The TDE models from \cite{Mockler19} show a similar increase in temperature at later times. \cite{Lodato11} demonstrate that for an opaque radiatively-driven wind, photons are released at a photospheric radius much larger than the launching radius, and as the accretion rate decreases, the photosphere sinks in, and the corresponding temperature increases. A good example of this process might be the TDE ASASSN-14ae \citep{Holoien14}, which shows a temperature evolution that resembles that of AT\,2018hyz, shown in Figure~\ref{fig:bolometric}.

\begin{figure}
	\begin{center}
		\includegraphics[width=\columnwidth]{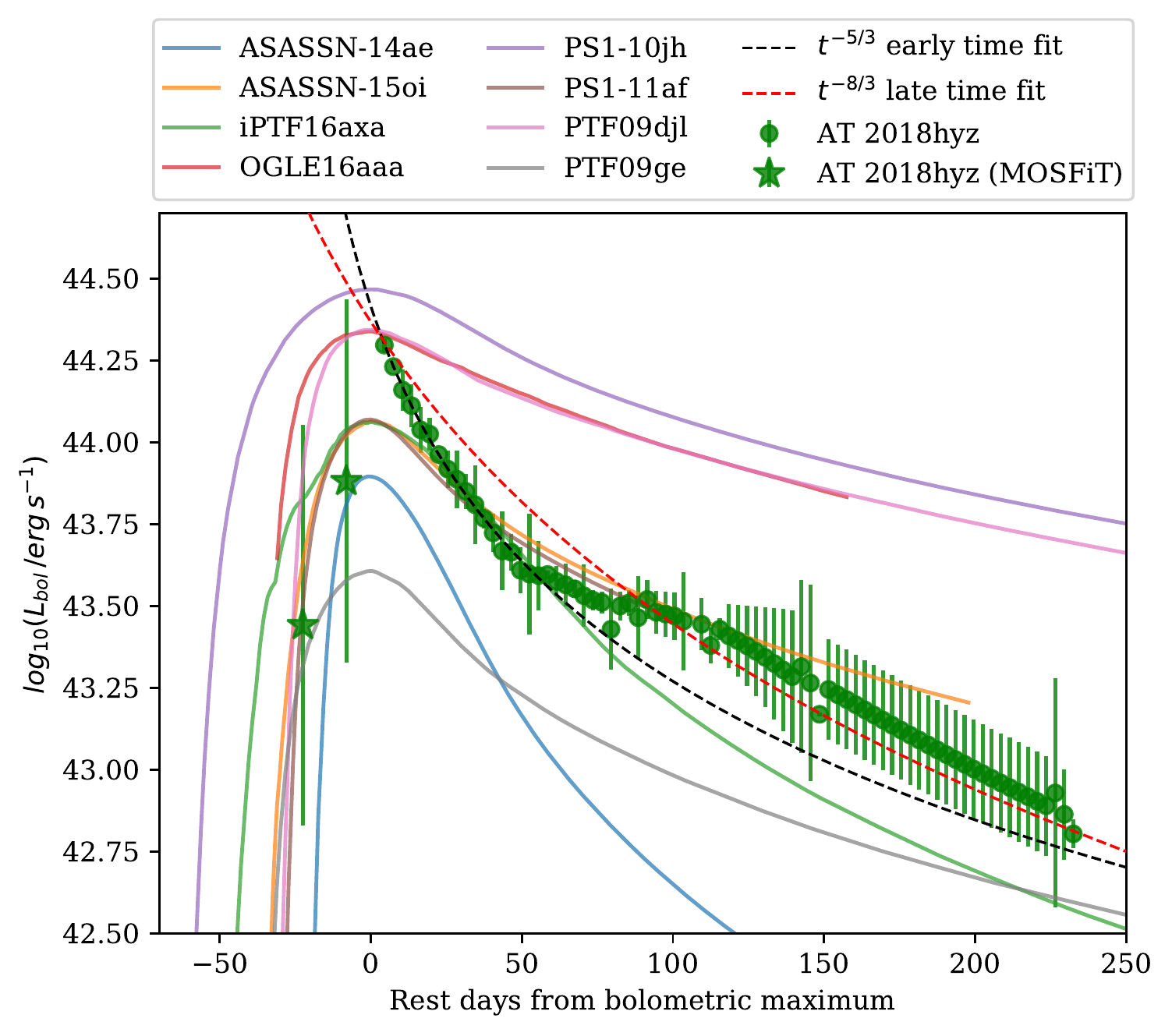}
		\caption{The bolometric light curve of AT\,2018hyz is shown in green. The corresponding bolometric light curve models of all other well-observed TDEs from the \citet{Mockler19} sample are shown for comparison. The black dashed line is a $t^{-5/3}$ fit to the early time data (phase $< 50$ days) of AT\,2018hyz, showing a clear bump in the late time light curve. The red dashes line shows a $t^{-8/3}$ fit to the late time data (phase $> 90$ days), the expected decline rate for a partial disruption \citep{Ryu20} \label{fig:comparisons}}
	\end{center}
\end{figure}

In Figure~\ref{fig:comparisons} we show the bolometric light curve of AT\,2018hyz as compared to other TDEs, and see that it is similar in luminosity and decline rate to some TDEs. The early time bolometric light curve (phase $<50$ days) is well fit by a power law that falls as $L \propto t^{-5/3}$, the theoretical decline rate expected for TDEs \citep{Rees88}. After a phase of 50 days the bolometric light curve deviates from a $t^{-5/3}$ decline, and we see a bump that lasts for $\gtrsim 100$ days, most pronounced in the UV. The TDE candidate ASASSN-15lh also shows a similar bump in the late time light curve, although much more pronounced than in AT\,2018hyz \citep{Leloudas16}. It should be noted that although ASASSN-15lh originated in the nucleus of a quiescent galaxy, it is a highly unusual event of uncertain nature, also suggested to be a superluminous supernova \citep{Dong16}. Swift J1644+5734 is another TDE that shows a bump in its light curve 30-50 days after peak, most prominent in bluer bands, same as for AT\,2018hyz. AT\,2018fyk also has a secondary optical bump \citep{Wevers19}, where the authors suggest that the second bump might be powered by efficient reprocessing of X-rays from a variable super-Eddington disk wind. One possible explanation for the bump in the bolometric light curve of AT\,2018hyz is the mechanism outlined in \cite{Leloudas16}. Those authors suggest that the light curves of TDEs are powered by two mechanisms: circularization of the debris, and accretion onto the black hole. For the smaller supermassive black holes, it is hard to disentangle these two; but for the most massive black holes ($\gtrsim 10^7$ M$_\odot$), the accretion disk will be thin, increasing the viscous timescale, allowing accretion to be observed in the form of a secondary peak in the light curve.

While only a few TDEs show a resolved secondary bump, many have now shown a flattening on timescales of hundreds of days after disruption; both at low black hole mass $\lesssim 10^{6.5}$ M$_\odot$ \citep{Velzen19} and at more typical masses, for example ASASSN-14li ($M_{\rm BH} = 10^{6.7}$ M$_\odot$; \citealt{Brown17}), ASASSN-18pg ($M_{\rm BH} = 10^{7.0}$ M$_\odot$; \citealt{Holoien20}), and AT\,2018zr ($M_{\rm BH} = 10^{6.9}$ M$_\odot$; \citealt{Velzen19_2018zr}). This flattening is likely due to an additional contribution of emission from the accretion disk. AT2018hyz shows both a secondary bump at around 50 days, and a flattening of the light curve at $\sim250-550$ days. If the late-time flattening indicates the formation of a thick accretion disk, we may need another mechanism to account for the earlier bump.

Instead, the flattening in AT\,2018hyz could be produced by a sudden outflow of material; supported by the fact that the time the flattening beings coincides with the appearance of two spectral lines blueshifted from H$\alpha$ and H$\beta$ by \mbox{$\sim 12,000$ km s$^{-1}$}, respectively \citep{Short20}. Additionally, the rise in temperature observed in AT\,2018hyz corresponds to the emergence of He II lines in the spectra, which develop after a phase of $\sim 70$ days and were suggested to be related to an outflow or material or colliding debris \citep{Short20}. If this is the case for AT\,2018hyz, it would be late compared to outflows launched from other TDEs, such as AT\,2018fyk, which showed a plateau 40 days after discovery, explained by either stream-stream collisions or subsequent accretion after the main peak \citep{Wevers19}. For comparison, ASASSN-14li showed an outflow which was estimated to be launched 20-30 days before its bolometric peak \citep{Alexander16}.

Adopting a total disrupted mass of \mbox{$0.01$ M$_\odot$} and a photospheric blackbody radius of $1.25\times10^{15}$ cm during the light curve peak (Figure \ref{fig:bolometric}), we measure the average density of material behind the photosphere to be \mbox{$\rho \approx 5\times10^{-15}$ g cm$^{-3}$}, which implies an optical depth $\tau \approx 0.8$ (for an opacity dominated by electron scattering in ionized hydrogen $\kappa = 0.34$ cm$^{2}$ g$^{-1}$). At later times, after the photosphere contracts to $3\times10^{14}$ cm, the corresponding optical depth is $\tau \approx 18$. The low optical depth at early times might allow us to peer deep into the TDE, allowing us to see disk signatures (double-peaked Balmer emission lines) more clearly than in other TDEs \citep{Short20}. The increasing optical depth may help to explain why we no longer see disk-like line profiles beyond $\sim 100$ days.

As noted, the UV brightness in a late-time \textit{Swift} observation taken at a phase of $\sim 550$ days suggests a relatively flat light curve between $\sim250-550$ days. The measured magnitude is brighter than we would expect from the {\tt MOSFiT} model shown in Figure~\ref{fig:mosfit}, and the expected model host photometry, listed in Table~\ref{tab:photometry}. Given that we know AT\,2018hyz to harbor an accretion disk from the spectroscopic observations of \cite{Short20}, the fact that we see a flat UV light curve at late times supports the interpretation of \cite{Velzen19} that late-time UV emission observed in TDEs is produced by long-term accretion. The integrated UV luminosity at this phase is \mbox{$L_{\rm >200nm} \approx 1.5\times10^{42}$ erg s$^{-1}$}, which corresponds to an accretion rate $\dot{M} \approx 2.6 \times10^{-4}$ M$_\odot$ yr$^{-1}$ assuming an efficiency $\epsilon = 0.1$.

\begin{table}
\caption{We obtained one epoch of \textit{Swift} photometry at a phase of $\sim 550$ days. The measured magnitude is brighter than the prediction from both the {\tt MOSFiT} TDE model and the pre-explosion model host photometry. The photometry is corrected for galactic extinction and has the host contribution subtracted.}
\label{tab:photometry}
\begin{tabular}{cccc}
\hline
\textit{Swift} Band & Photometry       & {\tt MOSFiT} Model & Host Model  \\
                    &  (542d)          &                    &             \\
\hline
UVW2                & $19.84 \pm 0.12$ & $21.28 \pm 0.04$   & 21.76       \\
UVM2                & $20.24 \pm 0.14$ & $21.27 \pm 0.04$   & 21.31       \\
UVW1                & $20.15 \pm 0.17$ & $21.34 \pm 0.04$   & 20.73       \\
U                   & $20.64 \pm 0.15$ & $21.60 \pm 0.05$   & 19.12       \\
\hline
\end{tabular}
\end{table}

\section{Conclusions and Discussion} \label{sec:conclusion}

AT\,2018hyz is a tidal disruption event found in the nucleus of a quiescent E+A galaxy at a redshift of $z = 0.04573$. We presented optical and UV photometry of AT\,2018hyz from UVOT, \mbox{ASAS-SN}, FLWO, and Las Cumbres, representing one of the best sampled TDE light curves in the literature (in addition to densely sampled spectroscopic observations, presented in \cite{Short20}), allowing us to study its evolution in detail. 

We modeled the light curves using {\tt MOSFiT} and find a best fit for a $5.2\times10^6$ M$_\odot$ black hole disrupting a $0.1$ M$_\odot$ star. Comparing to other similar {\tt MOSFiT} models of TDEs we find AT\,2018hyz to have the least disrupted star compared to the \cite{Mockler19} sample, with an impact parameter of just $\beta = 0.61$. This corresponds to an inferred total disrupted mass of $\lesssim 0.01$ M$_\odot$. A low disrupted mass may produce a low optical depth, which in turn allows us to see the accretion disk spectra with less reprocessing than other TDEs, which we observe in the form of double peaked hydrogen emission lines \citep{Short20}.

We detect a bump in the bolometric light curve after a phase of 50 days, most prominent in the UV, which could be due to a sudden outflow of material or reprocessing of X-rays into optical/UV radiation. This is consistent with the emergence of He II lines in the spectra and an increase in the bolometric temperature. We detect a strong UV-excess at a phase of $\sim 550$ days. Given that we know AT\,2018hyz has an accretion disk, evidenced by its spectra \citep{Short20}; this supports the interpretation of \cite{Velzen19} that suggests late-time UV excess in TDEs is produced by late-time accretion.

An X-ray source is detected up to a phase of 86 days, consistent with having a constant luminosity. The X-ray spectra can not be explained by a simple blackbody, but instead we find a best fit to an absorbed blackbody plus power-law model with a photon index of $\Gamma = 0.8\pm 0.6$. A non-thermal X-ray spectrum is expected for jetted TDEs. Extending the power-law component of the X-ray model to radio wavelengths predicts a radio flux in excess of our limits ($\lesssim 13.0\ \mu$Jy), thus the non-detection in the radio seems to favour inverse-Compton scattering for the non-thermal X-rays.

We consider three possible origins for the X-ray emission: the TDE itself, a pre-existing AGN, or star-formation. The latter can be excluded due to the negligible ongoing star-formation inferred from our radio data and host SED fitting. The X-rays could be consistent with a weak AGN given the measured luminosity, but temporal evolution in future deep X-ray observations will allow us to determine the nature of the X-ray emission.

The rich dataset we have presented and the finding of a very low disrupted mass indicates a new way to account for the diversity in observed TDEs.

\section*{Acknowledgements}
We thank B. Mockler for useful discussions regarding the {\tt MOSFiT} TDE model and an anonymous referee for comments towards the improvement of this paper. The Berger Time-Domain Group at Harvard is supported in part by NSF under grant AST-1714498 and by NASA under grant NNX15AE50G. S.~Gomez is partly supported by an NSF Graduate Research Fellowship. MN is supported by a Royal Astronomical Society Research Fellowship. KDA acknowledges support provided by NASA through the NASA Hubble Fellowship grant HST-HF2-51403.001 awarded by the Space Telescope Science Institute, which is operated by the Association of Universities for Research in Astronomy, Inc., for NASA, under contract NAS5-26555. IA is a CIFAR Azrieli Global Scholar in the Gravity and the Extreme Universe Program and acknowledges support from that program, from the Israel Science Foundation (grant numbers 2108/18 and 2752/19), from the United States - Israel Binational Science Foundation (BSF), and from the Israeli Council for Higher Education Alon Fellowship. Operation of the Pan-STARRS1 telescope is supported by the National Aeronautics and Space Administration under grant No. NNX12AR65G and grant No. NNX14AM74G issued through the NEO Observation Program. This work has made use of data from the European Space Agency (ESA) mission {\it Gaia} (\url{https://www.cosmos.esa.int/gaia}), processed by the {\it Gaia} Data Processing and Analysis Consortium (DPAC, \url{https://www.cosmos.esa.int/web/gaia/dpac/consortium}). Funding for the DPAC has been provided by national institutions, in particular the institutions participating in the {\it Gaia} Multilateral Agreement. This paper includes data gathered with the 6.5 meter Magellan Telescopes located at Las Campanas Observatory, Chile. Observations reported here were obtained at the MMT Observatory, a joint facility of the University of Arizona and the Smithsonian Institution. This research has made use of NASA's Astrophysics Data System. This research has made use of the SIMBAD database, operated at CDS, Strasbourg, France. Additional software used for this paper: Astropy\citep{astropy18},  PyRAF\citep{science12}, SAOImage DS9 \citep{Smithsonian00}, Matplotlib\citep{hunter07}, NumPy\citep{Oliphant07}, extinction \citep{Barbary16} PYPHOT(\url{https://github.com/mfouesneau/pyphot}).

\section*{Data Availability}

All the optical photometry used for this work and shown in Figure~\ref{fig:lightcurve} are available on the online supplementary material version of this article.

\bibliographystyle{mnras}
\bibliography{references}

\bsp
\label{lastpage}
\end{document}